%% file: Main.tex
\documentclass[dvipsnames,format=sigconf]{acmart}

\newcommand\doubleplus{+\kern-1.3ex+\kern0.8ex}

\AtBeginDocument{%
  }

\setcopyright{acmlicensed}
\copyrightyear{2018}
\acmYear{2018}
\acmDOI{XXXXXXX.XXXXXXX}

\acmConference[GECCO '24 Companion]{Genetic and Evolutionary Computation Conference}{July 14--18, 2024}{Melbourne, VIC, Australia}
\acmISBN{978-1-4503-XXXX-X/18/06}
\usepackage{lipsum}
\usepackage{subcaption}
\usepackage{bbm}
\usepackage{todonotes}

\newtheorem{dfn}{Definition}

\newcommand{\msat}{MAX-3SAT }

\newcommand{\tsat}{3SAT}
\newcommand{\sota}{state-of-the-art}

\usepackage[angle=0, color=red, pos={10px,30px},hanchor=l,firstpageonly=true, fontsize=0.35cm]{draftwatermark}
\SetWatermarkText{
  \shortstack{
       This is the author's version of the work (preprint). \\  
       The revised version will soon be published at the Genetic and Evolutionary Computation Conference (GECCO '24 Companion), July 14--18, 2024 \\
       https://doi.org/10.1145/3638530.3664153
  }
}

\begin{document}

\title{Using an Evolutionary Algorithm to Create (MAX)-3SAT QUBOs}

\author{Sebastian Zielinski}

\orcid{1234-5678-9012}
\affiliation{%
  \institution{Institute for Informatics, LMU Munich}
  \streetaddress{Oettingenstraße 67}
  \city{Munich}
  \country{Germany}
  \postcode{81545}
}
\email{sebastian.zielinski@ifi.lmu.de}

\author{Maximilian Zorn}
\affiliation{%
  \institution{Institute for Informatics, LMU Munich}
  \city{Munich}
  \country{Germany}
}
\email{maximilian.zorn@ifi.lmu.de}

\author{Thomas Gabor}
\affiliation{%
  \institution{Institute for Informatics, LMU Munich}
  \city{Munich}
  \country{Germany}}
\email{thomas.gabor@ifi.lmu.de}

\author{Sebastian Feld}
\affiliation{%
  \institution{Quantum \& Computer Engineering, Delft University of Technology}
  \city{Delft}
  \country{The Netherlands}}
  \email{s.feld@tudelft.nl}

\author{Claudia Linnhoff-Popien}
\affiliation{%
  \institution{Institute for Informatics, LMU Munich}
  \city{Munich}
  \country{Germany}}
\email{linnhoff@ifi.lmu.de}

\renewcommand{\shortauthors}{Zielinski et al.}

\input{Sections/01-Abstract}




\maketitle

\input{Sections/1-Introduction}
\input{Sections/2-Foundations}

\input{Sections/3-RelatedWork}
\input{Sections/4-Method}
\input{Sections/5-Experiments}

\input{Sections/6-Conclusion}

\input{Sections/9-Acknowledgements}

\newpage
\bibliographystyle{ACM-Reference-Format}
\bibliography{Main}

\end{document}

%% file: Sections/01-Abstract.tex
\begin{abstract}
 A common way of solving satisfiability instances with quantum methods is to transform these instances into instances of QUBO, which in itself is a potentially difficult and expensive task. State-of-the-art transformations from \msat to QUBO currently work by mapping clauses of a \tsat{} formula associated with the \msat instance to an instance of QUBO and combining the resulting QUBOs into a single QUBO instance representing the whole \msat instance. As creating these transformations is currently done manually or via exhaustive search methods and, therefore, algorithmically inefficient, we see potential for including search-based optimization. In this paper, we propose two methods of using evolutionary algorithms to automatically create QUBO representations of \msat problems. We evaluate our created QUBOs on 500 and 1000-clause \tsat{} formulae and find competitive performance to \sota{} baselines when using both classical and quantum annealing solvers. 
\end{abstract}

%% file: Sections/1-Introduction.tex
\section{Introduction}
Over the past decade, Quadratic Unconstrained Binary Optimization (QUBO) has become a unifying framework for modeling combinatorial optimization problems \cite{verma2021efficient}. A significant portion of the growing interest in QUBO can be attributed to the developments in the field of quantum computing, as QUBO serves as an input format for quantum annealers and the quantum approximate optimization algorithm on gate-based quantum computers. As a consequence of the increased availability and problem-solving capabilities of quantum hardware systems, researchers are investigating methods of transforming problems from various domains to instances of QUBO, to be able to evaluate the problem-solving capabilities of contemporary quantum hardware systems for their domain-specific problems. Examples of this effort include applications in finance, \cite{venturelli2019reverse,lang2022strategic}, logistics \cite{irie2019quantum,feld2019hybrid}, scheduling \cite{ikeda2019application,venturelli2015quantum}, and many more \cite{lucas2014ising}.

In this paper we are concerned with creating new methods for finding QUBO formulations for satisfiability problems. Given a formula of propositional logic, the satisfiability problem (SAT) asks, whether an assignment of Boolean values to the variables of the given formula exists such that the formula is satisfied. Satisfiability problems are ubiquitous in computer science. In theoretical computer science, they are often used to prove the NP-hardness of other problems. In practice, they occur in many application domains like circuit design and verification \cite{marques2000boolean}, dependency resolution \cite{abate2020dependency}, and planning \cite{rintanen2012planning}. As every satisfiability problem can efficiently be reduced to the case where each clause of the satisfiability problem consists of at most three variables (\tsat{}), we will focus on \tsat{} problems in this paper. The optimization version of a \tsat{} problem is a MAX-3SAT problem. In a MAX-3SAT problem, the goal is to find an assignment to the variables of a \tsat{} formula such that as many clauses as possible are satisfied. To transform MAX-3SAT instances into instances of QUBO, researchers developed several different methods (e.g., \cite{choi2010adiabatic,chancellor2016direct,nusslein2023solving}). It has been shown that the choice of a QUBO transformation for a given \msat instance can significantly impact the quality of the solutions when solving these problems on a quantum annealer \cite{zielinski2023influence,kruger2020quantum}.  Thus, creating QUBO formulations that yield better results when solving \msat problems on a quantum annealer can be seen as an effort to bring this widely studied method closer to practical application.

In this paper, we propose two methods for using evolutionary algorithms to create QUBO formulations for \msat problems. To the best of our knowledge, evolutionary algorithms have not yet been used to create QUBO formulations for specific instances of combinatorial optimization problems. One of the reasons for this is that it is potentially hard to create QUBO instances whose minima directly correspond to correct solutions of the given isntance of a combinatorial optimization problem without knowing any solution to the combinatorial problem in the first place.
For \msat problems, however, there is a problem decomposition method that creates such QUBO representations without having to know any solution to the \msat problem. This method consists of two steps. Let $\phi$ be a \tsat{} formula, containing $m$ clauses. In the first step, each of the $m$ clauses of $\phi$ will be transformed into an instance of QUBO. Every QUBO representation of a clause is constructed such that all satisfying assignments of a clause correspond to the minima of the QUBO representation of the clause. In contrast, the single non-satisfying assignment of a clause should have a higher (i.e., less optimal, since QUBO is a minimization problem) energy in the QUBO. This way, $m$ QUBO matrices are created (one QUBO matrix for each of the $m$ clauses). In the second step, the $m$ QUBO matrices corresponding to the $m$ clauses of $ \phi$ get combined into a single QUBO instance. Finding the minimum of this combined QUBO instance is equivalent to finding the optimal solution for a given \tsat{} problem.
Our contributions in this paper are:
\begin{enumerate}
    \item[1.] As previously described, to create a QUBO representation of a \msat problem, it suffices to transform each of its clauses into an instance of QUBO and then combine all resulting QUBO instances. Currently, two methods of finding transformations from \tsat{} clauses to QUBO exist. The first one is that a scientist manually creates this transformation; the second method uses the Pattern QUBO approach \cite{zielinski2023pattern}, which employs an exhaustive search. 
    
    As our first contribution, we propose an evolutionary algorithm that automatically finds QUBO representations for \tsat{} clauses, which is made feasible by the relatively small size of the Pattern QUBO matrices. Furthermore, we show how to adapt this method to create QUBO representations for \tsat{} clauses that fulfill certain design criteria (i.e., a desired sparsity, a specific gap between the optimal energies and the energies of incorrect solutions, $\dots$). Thus, we enable the automatic generation of specific QUBO representations for \tsat{} clauses without performing any procedure based on an exhaustive search nor a manual creation process.

    \item[2.] In current \sota{} methods of transforming \msat instances to instances of QUBO, each clause of the \tsat{} formula associated with the \msat problem is assigned one of the \emph{clause types} $\{0,1,2,3\}$. The clause type is the count of negated variables within the clause (e.g., a clause of type 0 possesses zero negated variables). Current \sota{} transformations then provide a single method of transforming each clause of a given type ($\{0,1,2,3\}$) to an instance of QUBO. Thus, all clauses of a given type (e.g., all clauses with zero negated variables) will be transformed into an instance of QUBO according to a single transformation rule.
    
    As our second contribution, we propose an evolutionary algorithm that can choose an individual QUBO transformation rule for each of the clauses of a \tsat{} formula, instead of using a single transformation rule for all clauses of the same type. As the solution landscape of a QUBO representation of a \msat problem is induced by the choices of the QUBO transformations for the \tsat{} clauses associated with the \msat problem, our approach enables the creation of QUBO representations with many different solution landscapes. The goal of our second evolutionary method is thus to find choices of QUBO transformation rules for clauses of a \tsat{} formula such that the resulting solution landscape is beneficial for a given optimizer.
\end{enumerate}

\noindent The remainder of this paper is organized as follows: In Sec.~\ref{sec:foundations}, we introduce the necessary foundations of satisfiability problems as well as population-based optimization and give a detailed explanation of the Pattern QUBO creation process. In Sec.~\ref{sec:relatedwork} we discuss related work on satisfiability as well as thematically related quantum optimization. We then formalize our evolutionary approach in Sec.~\ref{sec:method}, for the QUBO search and selection respectively, and discuss our experimental evaluation in the same order in Sec.~\ref{sec:experiments}. Finally, we conclude our findings in Sec.~\ref{sec:conclusion}.  

%% file: Sections/2-Foundations.tex
\section{Foundations}\label{sec:foundations}
\subsection{Satisfiability Problems}\label{subset:sat}
Satisfiability problems are concerned with the satisfiability of Boolean formulae. Thus, we first define a Boolean formula:
\begin{dfn}[Boolean formula \cite{arora2009computational}]
Let $x_1,..., x_n$ be Boolean variables. A \emph{Boolean formula} consists of the variables $x_1, ..., x_n$ and the logical operators  $\wedge$,  $\vee$,  $\neg$. Let $z \in \{0,1\}^n$ be a vector of Boolean values. We identify the value 1 as TRUE and the value 0 as FALSE. The vector $z$ is also called an \emph{assignment} as it assigns truth values to the Boolean variables $x_1, ..., x_n$ as follows: $x_i := z_i$, where $z_i$ is the $i$th component of $z$. If $\phi$ is a Boolean formula and $z \in \{0,1\}^n$ is an assignment, then $\phi(z)$ is the evaluation of $\phi$ when the variable $x_i$ is assigned the Boolean value $z_i$. If there exists a $z \in \{0,1\}^n$ such that $\phi(z)$ is TRUE, we call $\phi$ satisfiable. Otherwise, we call $\phi$ unsatisfiable.
\end{dfn}
Satisfiability problems are often given in conjunctive normal form, which we  define next:
\begin{dfn}[Conjunctive Normal Form \cite{arora2009computational}] A Boolean formula over variables $x_1,...,x_n$ is in \emph{Conjunctive Normal Form (CNF)} if it is of the following structure:
$$\bigwedge_i \big(\bigvee_{j} y_{i_j}\big)$$
Each $y_{i_j}$ is either a variable $x_k$ or its negation $\neg x_k$. The $y_{i_j}$ are called the \emph{literals} of the formula. The terms $(\vee_j y_{i_j})$ are called the \emph{clauses} of the formula. A \emph{$k$CNF} is a CNF formula, in which all clauses contain at most $k$ literals. 
\end{dfn}
Given a Boolean formula $\phi$ in \emph{$k$CNF}, the satisfiability problem is the task of determining whether $\phi$ is satisfiable or not. This problem was one of the first problems for which NP-completeness has been shown \cite{cook1971complexity}. In this paper, we will especially consider \emph{$3$CNF} problems, which we will refer to as \tsat{} problems.

The optimization version of a satisfiability problem is the MAX-SAT problem. In the MAX-SAT problem, we are given a Boolean formula $\phi$ consisting of $m$ clauses. The task is to find an assignment of truth values to the variables of $\phi$ such that as many clauses as possible are satisfied. Finding an assignment in the MAX-SAT problem that satisfies $m$ clauses is thus equivalent to solving the corresponding satisfiability problem (i.e., deciding whether $\phi$ is satisfiable or not). MAX-SAT is thus NP-hard as well. To avoid confusion, we want to emphasize that a MAX-3SAT instance consists of a \tsat{} formula (and not a ``MAX-3SAT formula,'' which does not exist as a notion), for which the maximum number of satisfiable clauses should be determined.

\subsection{Quadratic Unconstrained Binary Optimization (QUBO)}\label{subsec:qubo}
In this section we will formally introduce quadratic unconstrained binary optimization (QUBO) and related terminology that will be used in the remainder of this paper.
\begin{dfn}[QUBO~\cite{glover2018tutorial}] Let $\mathcal{Q} \in \mathbb{R}^{n \times n}$ be a square matrix and let $x \in \{0,1\}^n$ be an $n$-dimensional vector of Boolean variables. The QUBO problem is defined as follows:
\begin{equation}\label{eq:qubo}
 \text{minimize} \quad H_{\textit{QUBO}}(x) = x^T\mathcal{Q}x = 
\sum_{i}\mathcal{Q}_{ii}x_i + \sum_{i <j}\mathcal{Q}_{ij}x_ix_j
\end{equation}
\end{dfn}

 We call $H_{\textit{QUBO}}(x)$ the (QUBO) energy of vector $x$. The matrix $Q$ will also be called \emph{QUBO matrix}. Representing a QUBO matrix as an upper triangular matrix is customary.
In this paper, we will create QUBO instances that contain auxiliary variables. That means that, to transform a given \msat instance to an instance of QUBO, we introduce additional variables that do not correspond to any variables of the given \msat instance. We will often say that an assignment $\vec{x} = (x_1 := v_0, \dots, x_n := v_n), v_i \in \{0,1\}$ of Boolean values to the variables $x_1, \dots, x_n$ of the \msat instance has energy $E$ in $Q$, by which we mean:
\begin{equation}
    min\; \{(\vec{x},y)^T Q (\vec{x},y) \;|\; y \in \{0,1\}^m\} = E
\end{equation}
Here $Q$ is a QUBO matrix and $(\vec{x},y)$ is an  $(n+m)$-dimensional column vector defined as $(\vec{x},y) = (x_1 = v_0, \dots, x_n = v_n,y_1, \dots, y_m)$. The first $n$ values of the vector $(\vec{x},y)$ are given by the assignment $\vec{x} = (x_1 := v_0, \dots, x_n := v_n), v_i \in \{0,1\}$ of Boolean values to the variables $x_1, \dots, x_n$ of the \msat instance. The last $m$ entries represent the values of the auxiliary variables $y_1, \dots, y_m$.

\subsection{Pattern QUBOs}\label{subsec:patternqubos}

Let $\phi$ be a \tsat{} formula consisting of $m$ clauses. For each clause, we sort the variables such that all negated variables are always at the end of the clause. For example, the sorted version of the clause $(x_1 \vee \neg x_2 \vee x_3)$ is the clause ($x_1 \vee x_3 \vee \neg x_2)$. As a consequence of this sorting procedure, there are now only four different types of \tsat{} clauses in a \tsat{} instance:
\begin{enumerate}
    \item[] Type 0 --- no negations: $(a \vee b \vee c)$
    \item[] Type 1 --- one negation: $(a \vee b \vee \neg c)$ 
    \item[] Type 2 --- two negations: $(a \vee \neg b \vee \neg c)$ 
    \item[] Type 3 --- three negations: $(\neg a \vee \neg b \vee \neg c)$ 
\end{enumerate}

The idea of Pattern QUBOs is to find QUBO representations for each of these four types of clauses such that assignments that satisfy the clause correspond to the minimum in the QUBO representation. Likewise, assignments that do not satisfy a clause (of which there is only one for each clause) correspond to a non-minimal higher value in the QUBO representation. We apply the following procedure to transform any clause of $\phi$ to an instance of QUBO. First, we assess the type of the respective clause (i.e., we count the number of negated variables). Next, we sort the clause so that all negated variables are at the end of the clause. Finally, we replace all the variables of the Pattern QUBO of the same type as our current clause with the variables of the current clause (hence the name ``Pattern QUBOs''). After applying this procedure for each of the $m$ clauses of $\phi$ we combine the resulting $m$ QUBO representations of the clauses into a single QUBO representation.

To illustrate this procedure, suppose we are given the formula $\phi = (x_1 \vee x_2 \vee x_3) \wedge (x_1 \vee x_2 \vee x_4)$ and the Pattern QUBO for a type 0 clause shown in Tab.~\ref{tab:type_0_qubo}.
\begin{table}[!htb]
   \caption{Pattern QUBO for a type 0 clause $(a \vee b \vee c)$}
      \vspace{-.25cm}
\centering
\begin{tabular}[h]{|r||p{0.3cm}|p{0.3cm}|p{0.3cm}|c|}
\hline
& a & b & c & $y$ \\
\hline
\hline
a & & 2 & & -2 \\
\hline
b & & & & -2 \\
\hline
c & & & -1 & 1 \\
\hline
$y$ & & & & 1 \\
\hline
\end{tabular}
\label{tab:type_0_qubo}
\end{table}

The variable $y$ in the QUBO shown in Tab. \ref{tab:type_0_qubo} is an auxiliary variable that is needed to be able to guarantee that each satisfying assignment of the clause $(a \vee b \vee c)$ has minimum energy in the QUBO, while each non-satisfying assignment (which is only $a=b=c=0$) has a higher energy in the QUBO (see Sec. \ref{subsec:qubo}). To transform $\phi$ into an instance of QUBO we apply the procedure described above for each of the clauses of $\phi$. First, we observe that both clauses of $\phi$ do not contain any negations. Hence, both clauses are of type 0. As there are no negations, we do not have to sort the variables of any of the clauses. As our next step, we replace the variables of the Pattern QUBO shown in Tab. \ref{tab:type_0_qubo} with the variables of the first (or second, respectively) clause of $\phi$. The result of this step is shown in Tab. \ref{tab:pq_example}. 
\begin{table}[!htb]
   \caption{Applied Pattern QUBOs for given formula $\phi$}
    \begin{minipage}{.5\linewidth}
      \caption*{(a) $( x_1 \lor x_2\lor x_3 )$}
      \vspace{-.25cm}
\centering
\begin{tabular}[h]{|r||p{0.3cm}|p{0.3cm}|p{0.3cm}|c|}
\hline
& $x_1$ & $x_2$ & $x_3$ & $y_1$ \\
\hline
\hline
$x_1$ & & 2 & & -2 \\
\hline
$x_2$ & & & & -2 \\
\hline
$x_3$ & & & -1 & 1 \\
\hline
$y_1$ & & & & 1 \\
\hline
\end{tabular}
    \end{minipage}%
    \begin{minipage}{.5\linewidth}
      \centering
        \caption*{(b) $( x_1 \lor x_2 \lor  x_4 )$}
        \vspace{-.25cm}
\begin{tabular}[h]{|r||p{0.3cm}|p{0.3cm}|p{0.3cm}|c|}
\hline
& $x_1$ & $x_2$ & $x_4$ & $y_2$ \\
\hline
\hline
$x_1$ & & 2 & & -2 \\
\hline
$x_2$ & & & & -2 \\
\hline
$x_4$ & & & -1 & 1 \\
\hline
$y_1$ & & & & 1 \\
\hline
\end{tabular}
    \end{minipage} 
    \label{tab:pq_example}
\end{table}

In our last step, we now combine the two QUBO matrices, shown in Tab. \ref{tab:pq_example}, to receive a QUBO representation of $\phi$.

Note that, when minimizing a QUBO problem with QUBO matrix $Q$, we want to find the minimum of $\sum_{i}\mathcal{Q}_{ii}x_i + \sum_{i <j}\mathcal{Q}_{ij}x_ix_j$ (see Eq. \ref{eq:qubo}). Thus, in the case of the QUBO shown in Tab. \ref{tab:pq_example}(a) we want to find an assignment of Boolean values to the variables $x_1, x_2, x_3, y_1$  such that $P_1 = -x_3 +y_1 + 2x_1x_2 -2x_1y_1 -x_2y_2 +x_3y_1$ is minimized. Equivalently, in the case of the QUBO shown in Tab. \ref{tab:pq_example}(b) we want to find an assignment of Boolean values to the variables $x_1, x_2, x_4, y_2$  such that $P_2 = -x_4 +y_2 + 2x_1x_2 -2x_1y_2 -x_2y_2 +x_3y_2$ is minimized. As our last step, we add the polynomials $P_1$ and $P_2$ to receive the polynomial $P_3 = P_1 + P_2$. By minimizing $P_3$, we thus find an assignment of Boolean values to the variables $x_1, x_2, x_3, x_3, y_1, y_2$ that satisfies the formula $\phi$. Thus, by arranging the coefficients of $P_3$ in a matrix, we receive a QUBO matrix representation that, when minimized, yields optimal assignments for $\phi$.

%
%
\subsection{Population-Based Optimization}\label{subsec:population_optimization}
Let $\mathcal{X}$ be a state space. Let $\mathcal{T}$ be a target space with total order $\leq$. Let $\tau: \mathcal{X} \to \mathcal{T}$ be a target function. A tuple $\mathcal{E} = (\mathcal{X}, \mathcal{T}, \tau, E, \langle X_t \rangle_{0\leq t \leq g})$ is a population-based optimization process iff $X_t \subseteq \mathcal{X}$ for all $t$ and $E$ is a possibly randomized or non-deterministic function so that the population-based optimization run is produced by calling $E$ repeatedly, i.e., $X_{t+1} = E(\langle X_u \rangle_{0 \leq u \leq t}, \tau)$ where $X_0$ is given externally or chosen randomly. \cite{gabor2021productive}

Let $\mathcal{E} = (\mathcal{X}, \mathcal{T}, \tau, E, \langle X_u \rangle_{0 \leq u \leq t})$ be a population-based optimization process. One realization of population-based optimization is the concept of an \textit{Evolutionary Algorithm} (EA). The process $\mathcal{E}$ continues via such an evolutionary algorithm if $E$ has the form $$E(\langle X_u \rangle_{0 \leq u \leq t}, \tau) = X_{t+1} = \textit{selection}\big(X_t \cup \textit{variation}(X_t)\big)$$ where $\textit{selection}$ and $\textit{variation}$ are possibly randomized or possibly non-deterministic functions so that $|\textit{selection}(X)| \leq |X|$ and $|\textit{variation}(X)| \geq |X|$ and $\big| \textit{selection}\big(X \cup \textit{variation}(X)\big)\big| = |X|$. In the context of evolutionary search, the term \textit{generation} describes one such iteration of $X_t \to X_{t+1}$ and \textit{fitness} refers to the evaluation of an individual on the target-function, i.e., $\textit{fitness(x)} := \tau(x)$. The detailed expression of these operators will follow in Sec.~\ref{sec:method}.




%% file: Sections/3-RelatedWork.tex
\section{Related Work}\label{sec:relatedwork}
To express \msat problems as instances of QUBO, researchers have developed many different methods over the past decades. These methods include procedures that are based on polynomial-time reduction of \tsat{} problems to instances of maximum independent set \cite{choi2010adiabatic}, procedures that count satisfied clauses \cite{nusslein2023solving}, and many more \cite{nusslein2023solving,zielinski2023pattern,chancellor2016direct,verma2021efficient}. The most common method of transforming \msat instances to instances of QUBO is to transform each clause of the \msat instance to an instance of QUBO and then sum up all the resulting QUBO matrices, as demonstrated in Sec. \ref{subsec:qubo}. These approaches include methods representing a given clause as a pseudo-Boolean function and using auxiliary variables to reduce any higher-dimensional (i.e., non-quadratic) terms to quadratic terms. This can be achieved by quadratic reformulation techniques \cite{rosenberg1975reduction,anthony2017quadratic}. Furthermore, many approaches, like Chancellor's \cite{chancellor2016direct} and Nüßlein's \cite{nusslein2023solving} approach, transform each clause to an instance of QUBO according to some custom logic that leads to QUBO instances in which the following \textbf{energy condition} holds:
\begin{enumerate}
    \item[1.] If an assignment of Boolean values to the variables of the \tsat{} clause satisfies this clause, then this assignment should have minimal energy in this QUBO.
    \item[2.] If an assignment of Boolean values to the variables of the \tsat{} clause does not satisfy this clause, then it must have a higher, non-optimal energy in this QUBO.
\end{enumerate}

This approach introduces one auxiliary variable for each clause of the \msat instance. All of these methods have in common that a scientist manually crafted these QUBO transformations.

The Pattern QUBO method \cite{zielinski2023pattern} is an algorithmic procedure that can identify QUBO representations of \tsat{} clauses automatically (i.e., not manually). This procedure can be seen as a generalization of all approaches that transform clauses to instances of QUBO according to the just described logic. As in the case of Chancellor's and Nüßlein's transformation, the Pattern QUBO method introduces an additional auxiliary variable for each clause of the \msat instance. To transform an arbitrary clause of a \msat problem into an instance of QUBO, the Pattern QUBO method thus starts with an empty $4\times4$-dimensional QUBO matrix (three variables corresponding to the variables of the \tsat{} clause + one auxiliary variable). As QUBO matrices are commonly upper triangular matrices, a $4\times4$-dimensional QUBO matrix possesses 10 entries. The user next specifies a set of values (e.g., $\{-1,0,1\}$) that the Pattern QUBO procedure is allowed to insert into the QUBO matrix. Finally, an exhaustive search procedure tries all possible combinations of assigning values from the specified set of allowed values to the 10 entries of the QUBO matrix to find QUBO matrices that satisfy the above-mentioned \emph{energy condition}. Executing this procedure leads to multiple possibilities of transforming a given \tsat{} clause to an instance of QUBO. Choosing any of those possibilities for a given clause results in a mapping from a \tsat{} clause to an instance of QUBO. Combining all the QUBO instances (as demonstrated in Sec \ref{subsec:patternqubos}) that result from transforming \tsat{} clauses to instances of QUBO leads to a QUBO representation of the whole \msat problem.

In contrast to the methods explained above, the genetic algorithm method we propose in this paper is a method that can automatically create QUBO representations that fulfill certain criteria (i.e., a predefined sparsity, a given energy gap between incorrect answers and correct solutions) for arbitrary \tsat{} clauses. In contrast to the Pattern QUBO method, our genetic algorithm enables finding QUBO representations in much larger spaces, as the Pattern QUBO method uses an exhaustive search procedure that becomes increasingly infeasible the larger the set of possible values specified by the user becomes.

Furthermore, various forms of evolutionary optimization, like genetic algorithms (GA) \cite{Holland_1992_ga}, evolutionary algorithms (EA) \cite{Bäck_Schwefel_1993_ea}, memetic algorithms (MA) \cite{Neri_Cotta_2012_ma,Moscato_Cotta_Mendes_2004_ma}, and other forms of metaheuristic algorithms \cite{wang2022metaheuristic} have recently found application in quantum optimization problems: Both GAs and EAs have been used as classical optimizers for approximate quantum optimization \cite{Acampora_Chiatto_Vitiello_2023_a, Acampora_Chiatto_Vitiello_2023_b} and (random) Ising systems \cite{Prügel-Bennett_Shapiro_1997, Maksymowicz_Galletly_Magdon_Maksymowicz_1994}. Other studies utilized evolutionary strategies for the training of quantum classifiers, emphasizing the adaptability of EAs and GAs to quantum machine learning \cite{Chen_Huang_Hsing_Goan_Kao_2022, Friedrich_Maziero_2023}. Noisy quantum effects themselves have even been leveraged in the development of genetic algorithms, e.g., using quantum fluctuation for mutations of individuals \cite{King_Mohseni_Bernoudy_Fréchette_Sadeghi_Isakov_Neven_Amin_2019,gabor2022modifying}.

However, even as many of these approaches have leveraged quantum computing (and quantum annealing using the QUBO models) to optimize the solutions and formulation of QUBO models, literature on actual \textit{generation} of QUBO matrices is rather sparse. Quite recently, the AutoQUBO project has proposed ideas for constructing general-purpose optimization QUBOs algorithmically \cite{Moraglio_Georgescu_Sadowski_2022, Pauckert_Ayodele_García_Georgescu_Parizy_2023}, many of which share the concept of decomposing the optimization problem at hand, similar to the Pattern QUBO approach for \tsat{}. Similarly, some of the metaheuristic approaches described in the survey of \cite{wang2022metaheuristic} formulate ways to optimize (or decompose) the mathematical QUBO formulations for algorithmic composition of QUBOs, although with more focus on the (assignment-)solution quality rather than the purpose of creating principled QUBO structures like we do in this work.

%% file: Sections/4-Method.tex
\section{Method}\label{sec:method}
We will now detail the approach and parameterization of the population-based evolutionary search that we employ to firstly search correct, principled QUBO patterns and then secondly utilize the found pattern set(s) within an evolutionary selection of patterns for individual clauses in a \tsat{} formula. 

Our evolutionary buildings blocks \textit{mutate, recombine, select} and \textit{migrate}, serving the purpose of \textit{selection} and \textit{variation} as formalized in Sec.~\ref{subsec:population_optimization},  are realized as follows for the entirety of our work:

Let $\textit{one-point mutate} : \mathcal{X} \to \mathcal{X}$ be a randomized mutation function of a single individual (solution candidate) $x$ in the population $X$. We write $\textit{mutate}[\cdot]$ for the per-individual application of \textit{one-point mutate}$(\cdot)$, taking place with probability \textit{mut-rate} $\in [0;1)$. Should mutation occur, one point $i \in [0;|x|)$ from an individual's genome is sampled and the corresponding value-entry is replaced with a valid value-entry of the solution space $\mathcal{X}$. 

Let $\textit{one-point crossover} : \mathcal{X} \times \mathcal{X} \to \mathcal{X}$ be a randomized, two-parent recombination function for two parental individuals $x_1, x_2 \sim X$ that produce an offspring $x_3 \in \mathcal{X}$. The new individual combines the first $j$ parts of the solution candidate $x_1$, i.e, $x_{1, 0 \dots j}$, and the remaining part of the solution candidate $x_2$, i.e., s $x_{2, j+1 \dots |x_i|}$, where $j \sim [0;|x_i|)$ is randomly drawn for each parent pair, i.e., $x_3 =  x_{1, 0 \dots j} \doubleplus x_{2, j+1 \dots |x_2|}$. Recombined samples are drawn until the population has increased by a percentage of \textit{par-rate} $\in [0;1)$, which we fix to \textit{par-rate}=0.3 for this work. We write $\textit{recombine}[\cdot]$ for the recursive application of $\textit{one-point crossover}$, for two parental individuals $x_1, x_2 \sim X$ that each get randomly sampled (without replacement) from $X$ and chosen with probability \textit{rec-rate} $\in [0;1)$. 

Let $\sigma_N^{\textit{roulette}}: \wp(\mathcal{X}) \to \wp(\mathcal{X})$ be a randomized selection function that returns $N \in \mathbb{N}$ individuals, i.e., $|\sigma_N(X)| = N$ and $\sigma_N(X) \subseteq X$ for all $X$. In our case we employ roulette selection, a commonly used selection algorithm (cf. \cite{lipowski2012roulette, yu2016improved, zhang2012equal}) that selects individuals with a probability proportional to their fitness, i.e., individuals $x \sim X$ are drawn from the population (with replacement) and individual $x_i$ is selected with a probability of $ P(x_i) = \frac{\textit{fitness} (x_i)}{\sum_{x \in X} \textit{fitness(x)}}$. Upon selection, individual $x_i$ is removed from the population and the selection continues with $\sigma^{\textit{roulette}}(\{X \setminus x_i\})$. We also employ \textit{elitism}, where we directly include the best percent of the population (\textit{elt-rate} $\in [0;1)$), therefore at the end of each iteration we select $|\sigma_{|X|-|X|*\textit{elt-rate}}^{\textit{roulette}}(X)|$ non-elite individuals to remain in the population for the next generation.

\subsection{Evolutionary Pattern QUBO Creation}\label{subsec:method_search}
As explained in Sec.~\ref{sec:relatedwork}, a $4\times4$-dimensional QUBO matrix is needed to express a \tsat{} clause as an instance of QUBO. Hence a solution candidate $Q_C$ is an upper triangular $4\times4$ QUBO matrix. For our genetic algorithm, we will represent a solution candidate $Q_C$ as an array $[q_1, q_2, ..., q_{10}]$. Here the values $q_1, ..., q_4$ denote the first row of the upper triangular matrix of $Q_C$, the values $q_5, ..., q_7$ denote the second row of $Q_C$, etc. Each array entry can be filled with an arbitrary value from a user-specified predefined value range (e.g., the set $\{-1,0,1\}$). Let $Q_C$ be a solution candidate for a type $i$, $0 \leq i \leq 3$, clause. Let $S_{SAT}$ be the set of all satisfying assignments for a clause of type $i$. Let $E_{SAT}$ be the set of all energies of all the assignments that satisfy a clause of type $i$. Let $E_{UNSAT}$ be the energy of the assignment that does not satisfy the clause of type $i$. Finally let Energy(assignment) be the energy of an assignment, calculated as described in Sec. \ref{subsec:qubo}.

We compute the fitness of a solution candidate $Q_C$ via the following fitness functions:
\begin{itemize}
    \item \textbf{Uniformity:} All satisfying assignments of a clause of type $i$ should have the same energy.    Let $min\_correct := min\; E_{SAT}$. Penalize deviations with $-\sum_{a \in S_{SAT}}|min\_correct - Energy(a)|$.
    \item \textbf{Correctness:} 
    All satisfying assignments of a clause of type $i$ should have a lower (i.e., better) energy than the non-satisfying assignment.
    Let $S$ be the set of all correct assignments for a clause of type $i$ with a higher (i.e., worse) energy than the non-satisfying assignment for a clause of type $i$.
    Let $C$ be a positive constant.
    Penalize deviation from the desired outcome with $-\sum_{a \in S}|E_{UNSAT} - Energy(a)| * C + C$. 
   \item \textbf{Sparsity:} We want to enforce a specific sparsity $sp$. Penalize deviations with $- |sp - \sum_{i=0}^{10} \mathbb{I}[x_i \neq 0]|$, where $ \mathbb{I}$ is the indicator function of non-zero QUBO matrix entries.
    \item \textbf{Energetic Gap:}
    We want to enforce a certain gap between the energy of the correct solutions and the energy of the non-satisfying assignment. Let
    $max\_correct := max\; E_{SAT}$.
    Let $desired\_gap$ be an integer that represents the desired gap between the worst correct energy and the energy of the non-satisfying assignment.
    Penalize deviations with\\ $-||max\_correct - E_{UNSAT}| - desired\_gap|$
\end{itemize}

Our target function $\tau$, $\mathcal{T} \subseteq \mathbb{R}$, for our EA is then computed by accumulating penalties for combinations of the above fitness criteria. Individuals satisfying all criteria, therefore, have a fitness of $0$. 

\subsection{Evolutionary Pattern QUBO Selection}\label{subsec:method_selection}
As explained in Sec.~\ref{subsec:patternqubos}, to create a QUBO representation of a \msat instance, one transforms each clause of a \tsat{} formula associated with the \msat problem to an instance of QUBO and combines all the resulting QUBOs into one single QUBO instance. Currently known approaches propose a fixed method of transforming each clause \textbf{type} to an instance of QUBO. Each of these (four) mappings is then applied to the clauses of the \tsat{} formula of the correct type. In this section, we will introduce the idea of mapping each \textit{clause} (not clause type!) individually to a specific instance of a QUBO pattern and then combine the resulting QUBO selections to receive a transformation from a \msat problem to an instance of QUBO. We will use an evolutionary method to choose the individual QUBO mapping for each clause of the \tsat{} formula as follows. Let $m$ be the number of clauses of the \tsat{} instance. First, we are given four ordered sets of valid Pattern QUBOs $S_i$, $ i \in \{0,1,2,3\}$ --- one set for each clause type. An individual $x$ of the population $X$ is represented by a list of $m$ integers $x := [l_1, l_2, ..., l_m]$. Each of the integers $l_k$, $1 \leq k \leq m$, corresponds to an index of the correct Pattern QUBO set. Thus, if the $k$th clause is of type $t_k$ ($0 \leq t_k \leq 3$), then the $k$th entry of $x$ (which is $l_k$) denotes that, for the $k$th clause, the Pattern QUBO $l_k$ of the set $S_{t_k}$ should be used.



Upon evaluation, the $m$ Pattern QUBOs specified by the $m$ integers $l_1, ..., l_m$ and the clause types $t_1, ..., t_m$ of the individual $x$ are combined into a single QUBO instance (as described in Sec.~\ref{subsec:patternqubos}), corresponding to the \msat problem, and solved $N$ times using D-Wave's \textit{Advantage\_system6.4} quantum annealer \cite{dwaveQuantum} and D-Wave's tabu search algorithm \cite{dwaveTabu}. Either method returns an assignment $\vec{x} = (x_1 := v_0, \dots, x_n := v_n), v_i \in \{0,1\}$, of Boolean values to the variables $x_1, \dots, x_n$ of the \tsat{} formula that the current population is optimizing for. As the target function $\tau$, $\mathcal{T} \in \mathbb{R}$, --- and therefore the fitness of the individual --- we simply use the highest count of satisfied clauses across the $N$ assignment trials.

\subsection{Baselines}\label{subsec:method_baseliens}
In Sec. \ref{sec:experiments} we will perform a case study, in which we solve a set of \tsat{} formulas with different QUBO formulations on D-Wave's quantum annealer Advantage\_System6.4 and D-Wave's implementation of a tabu search method. The QUBO methods we will compare our result to are:
\begin{itemize}
    \item Chancellor's method, as described in the special case section of \cite{chancellor2016direct}. 
    \item Nüßlein's $n+m$ method, as described in \cite{nusslein2023solving}
    \item Random-Fixed-Pattern: Given a set of correct Pattern QUBOs for each of the four types of clauses (see Sec. \ref{subsec:patternqubos}), we choose a single Pattern QUBO for each of the four clause types at random and reuse these chosen Pattern QUBOs for all clauses of the respective types within a given \tsat{} formula.
    \item Random-Individual-Pattern: Given a set of correct Pattern QUBOs for each of the four types of clauses. For each individual clause of a given \tsat{} instance, we choose an arbitrary Pattern QUBO of the correct type at random.
\end{itemize}
Our final baseline is directly guessing random solutions for the given \tsat{} instances, without any transformations to QUBO. We will call this baseline simply \emph{random}.


%% file: Sections/5-Experiments.tex
\section{Experiments}\label{sec:experiments}
We now layout our experiments in two parts. We first apply the concepts of creating, i.e., searching, valid Pattern QUBOs via the evolutionary approach we have described in Sec.~\ref{subsec:method_search}. With a subset of the found Pattern QUBOs, we then proceed to select individual Pattern QUBOs for each \tsat{} clause of a \tsat{} formula according to Sec.~\ref{subsec:method_selection} and evaluate their performance in comparison to an ensemble of random baselines, as well as \sota{} QUBO transformations for the \msat problem. The `search' and `selection' aspects correspond to the proposed contributions 1 and 2 respectively.

\subsection{Evaluation EA QUBO Search}

\begin{figure*}[ht!]
    \centering
    \begin{subfigure}{\textwidth}
        \includegraphics[width=0.9\linewidth]{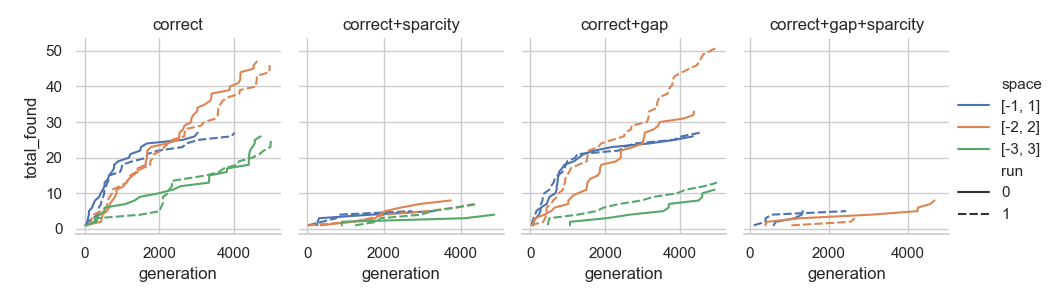}
        \caption{Pattern search over generations (small ranges)}
        \label{fig:patternsearch_small_line}
    \end{subfigure}//
    \begin{subfigure}{\textwidth}
    \centering
        \includegraphics[width=0.99\linewidth]{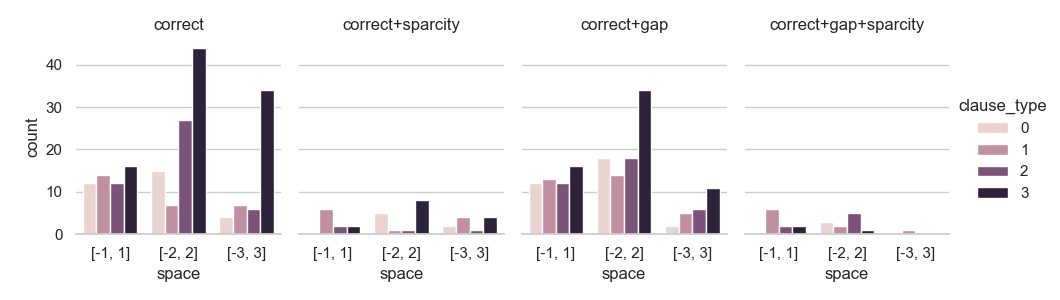}
        \caption{Count of found patterns (small ranges)}
        \label{fig:patternsearch_small_count}
    \end{subfigure}\\
    \begin{subfigure}{0.45\textwidth}
    \centering
        \includegraphics[width=0.95\linewidth]{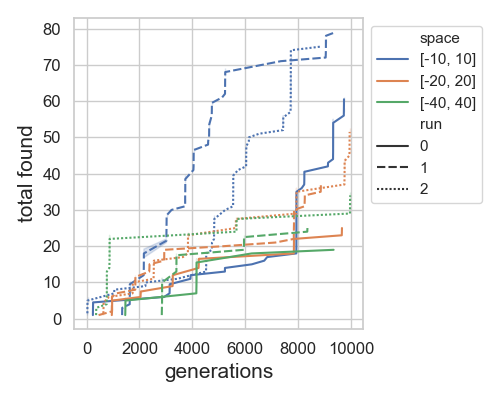}
        \caption{Pattern search over generations (big ranges)}
        \label{fig:patternsearch_big_line}
    \end{subfigure}%
    \begin{subfigure}{0.45\textwidth}
    \centering
       \includegraphics[width=0.95\linewidth]{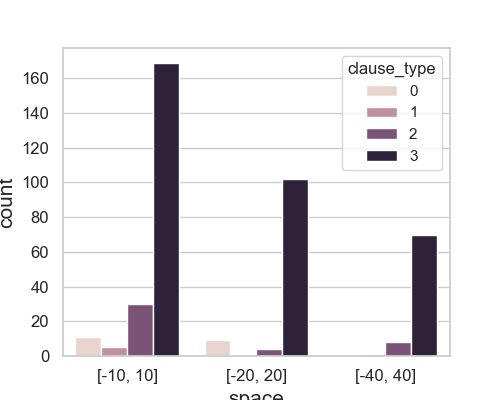}
       \caption{Count of found patterns (big ranges)}
       \label{fig:patternsearch_big_count}
    \end{subfigure}
    \caption{
        Frequency over time (a) and distribution of found pattern types (b) in evolutionary search across three smaller integer spaces: $[-1, 1], [-2, 2]$, and $[-3, 3]$ over 2 runs each. Comparisons are made between searches for correct patterns (b, left), patterns with an energy gap requirement (here gap:$1$, b, mid-left), patterns with a sparsity requirement (here sparsity:$7$, b, mid-right), and patterns satisfying both gap and sparsity constraints (b, right). Each run consists of $5000$ total generations ($500$ repetitions of fresh populations evolving over $10$ generations, $100$ individuals per population with \textit{mut-rate, rec-rate}$= 0.5$, \textit{elt-rate, mig-rate}$=0.1$).
        To show adaptability to larger ranges we repeat the experiment of finding correct patterns over time (c) and the corresponding count per pattern type (d) with three big spaces ($[-10, 10], [-20, 20]$, and $[-40, 40]$) as well. Each of the $3$ runs in (c,d) consists of $10,000$ total generations ($100$ repetitions of fresh populations evolving over $100$ generations, $300$ individuals per population with \textit{mut-rate, rec-rate}$= 0.5$, \textit{elt-rate, mig-rate}$=0.1$).
    }
    \label{fig:experiment_1}
\end{figure*}

We use the EA to search for correct\footnote{We refer to the criteria combination of \textit{Uniformity} + \textit{Correctness} as `correct' and `valid' interchangeably, since \textit{Uniformity} is a generally desired property.} Pattern QUBOs (with the criteria \textit{Uniformity}, \textit{Correctness}) as well as Pattern QUBOs that fulfill specific design criteria, like a specific energy gap (with the \textit{Energetic Gap} criteria, here gap=1), a predefined sparsity (with the \textit{Sparsity} criteria, here sparsity=$7$), as well as both of them together (with all four criteria), fulfilling both the gap and sparsity requirement. Fig.~\ref{fig:patternsearch_small_line} shows the results of the (integer-range) search spaces $[-1;1]$, $[-2;2]$ and $[-3;3]$. In contrast to satisfying the gap requirement, reaching valid solutions with the correct sparsity is more difficult (esp. for early-generations population), which may also be expected since there are simply fewer valid solutions under this constraint.

We observe that the amount of Pattern QUBOs found decreases as the search space broadens, which is to expected.\footnote{For an intuition, in the range $[-1;1]$ there are $27$ out of $3^{10}$ valid pattern-value combinations for the four clause types. While it is hard to efficiently determine exactly \textit{how} many correct combinations there are, the space increases drastically upon each range-increase, i.e., $[-2;2]: 5^{10}$ possibilities, $[-3;3]: 7^{10}$ possibilities, etc.} Evidently, the EA effectively and efficiently identifies more valid patterns in more constrained, smaller search spaces within approximately $1000$--$2000$ total generations. However, it struggles with larger spaces and additional constraints, highlighting the challenges in balancing search space breadth with constraint satisfaction.

Furthermore, Fig.~\ref{fig:patternsearch_small_count} shows that the found Pattern QUBOs are distributed fairly evenly between the four types of \tsat{} clauses. For larger ranges, this distribution becomes heavily skewed towards the type 2 and type 3 Pattern QUBOs. Interestingly, as shown in Fig.~\ref{fig:patternsearch_big_count}, this skewing effect is even more pronounced as we drastically increase the search space size to the ranges $[-10;10]$, $[-20;20]$, and $[-40;40]$. The question as to why the EA search seems to find such patterns more easily would require a larger-scale ablation study, which we leave for future work.

For the larger search spaces $[-10;10]$, $[-20;20]$, and $[-40;40]$, searching for valid individuals becomes drastically more difficult, which is why we abandon the `fail fast' idea and give the populations enough generations to converge to good solutions (i.e., correct patterns). Fig.~\ref{fig:patternsearch_big_line} gives an impression of the computational scalability of the evolutionary approach. Larger spaces still remain challenging, but the fact that patterns are found still makes for a promising alternative to brute-force search, which becomes increasingly infeasible as the search space grows. 

\subsection{Evaluation EA QUBO Selection}

\begin{figure*}[ht!]
    \centering
    \begin{subfigure}{0.33\textwidth}
        \includegraphics[width=.95\textwidth]{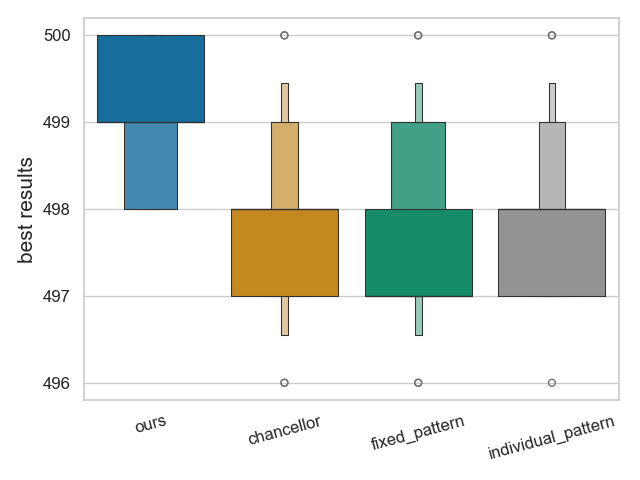}
        \caption{500-clause DWAVE tabu-search}
        \label{fig:tabu_dwave_500}
    \end{subfigure}%
    \begin{subfigure}{0.33\textwidth}
    \centering
        \includegraphics[width=.95\textwidth]{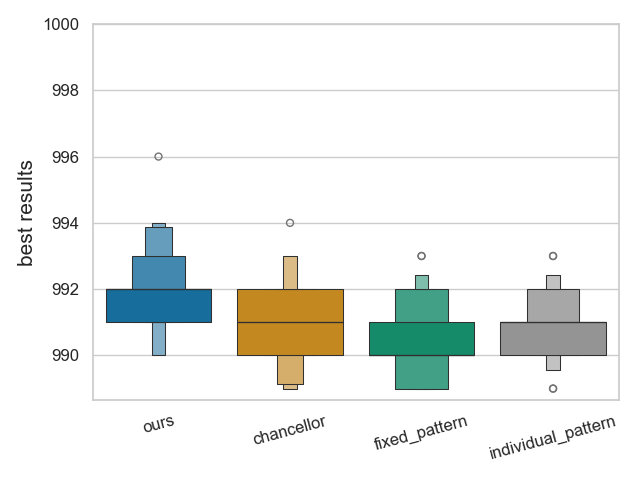}
        \caption{1000-clause DWAVE tabu-search}
        \label{fig:tabu_1000_hard}
    \end{subfigure}
    \begin{subfigure}{0.33\textwidth}
    \centering
      \includegraphics[width=.95\textwidth]{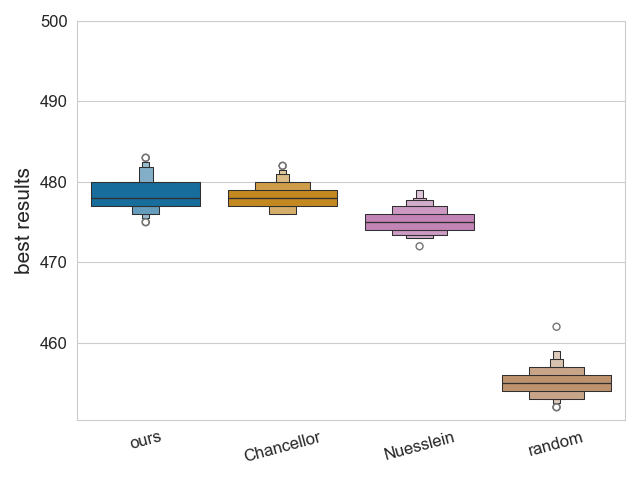}
      \caption{500-clause DWAVE \textit{Advantage\_system6.4}}
      \label{fig:quantum_dwave_500}
    \end{subfigure}%
    \caption{
       Evolutionary pattern selection across datasets with varying clause length and clause complexity. (a) Performance distribution on a 500-clause dataset with a standard difficulty ratio, showcasing the number of clauses satisfied by our EA (blue) in comparison to baseline methods \textit{chancellor} (gold), \textit{fixed-patterns} (teal) and \textit{individual-pattern} (grey). (b) Outcomes on a more complex 1000-clause dataset against the same baselines.
       (c) Performance of the EA on a 500-clause dataset using the D-Wave Advantage\_system6.4 quantum computer, compared to \textit{chancellor}, \textit{nüsslein} (violet) and \textit{random} guessing (brown) indicating the transfer-ability of EA optimizations from proxy solvers to actual quantum hardware. Results are obtained using parameters of 100 individuals over 50 generations, 5-tabu reads with 150ms timeout for the EA, and best results of $50*5$-tabu reads (also 150ms timeout) for all other baselines. 
    }
    \label{fig:experiment_2}
\end{figure*}

We then task the EA to optimize a selection of individual patterns for each clause in a formula from a predefined set per clause-type. Since all $27$ valid patterns for range $[-1,1]$ are known (comparatively quickly found by brute-force search), and we also have already found all of them in the previous experiment (Fig~\ref{fig:patternsearch_small_count}), we provide the EA with the following sets of available patterns to choose from: \{type 0: $6$, type 1: $7$, type 2: $6$, type 3: $8$\} ($27$ in total).

We construct two datasets to evaluate our approach on. The first one features $100$ problems of $500$ clauses each and $145$ variables (for the commonly used `difficult' clause-variable ratio of $4.2$ \cite{gent1994sat} where the clauses are sampled uniformly). For the second dataset we construct a set of non-uniformly sampled, empirically difficult $50$ formulae with $1000$ clauses each and $298$ variables according to the \textit{Balanced SAT} method of Spence \cite{spence2017balanced}.

We evaluate the EA (cf. Sec.~\ref{subsec:method_selection}) on the 500-dataset against the baselines \textit{chancellor}, \textit{random-individual-pattern}, and \textit{random-fixed-pattern} (cf. Subsec.~\ref{subsec:method_baseliens}). We find that our approach is able to find fully satisfying solutions (500/500 satisfied clauses) much more reliably across the $100$ formulae, in comparison to the baselines, where most of the solutions satisfy 2 fewer clauses. Fig.~\ref{fig:tabu_dwave_500} shows the boxen plots\footnote{See \url{https://vita.had.co.nz/papers/letter-value-plot.html}.} of this distribution, where our approach is able to find satisfying solutions for 34 out of 100 \tsat{} formulae, compared to the baselines which only find 2 satisfying solutions each out of a 100 \tsat{} formulae. 

Similarly, we evaluate the EA classically with tabu-search (5-tabu-reads with 150ms timeout each) on the 1000-dataset against the same baselines \textit{chancellor}, \textit{random-individual-pattern}, and \textit{random-fixed-pattern}, with the same parameters. Results can be seen in Fig.~\ref{fig:tabu_1000_hard}, and although the advantage is not quite as pronounced, the majority of answers is better than or similar in performance to \textit{chancellor}; we observe potential for a few very good solutions (up to 996/1000 solved clauses).

Finally, we evaluate the 500-clause dataset on real quantum hardware (\textit{D-Wave's Advantage\_system6.4}, with $10\times100$ shots), against the baselines \textit{Chancellor}, \textit{Nüsslein} and \textit{random} guessing. We find slightly advantageous performance in comparison to \textit{Chancellor}, which is interesting and hints at the transfer-ability of training evolutionary approaches on proxy-solvers like the tabu-search. We also directly outperform the \sota{}-approach of \textit{Nüsslein} as well as the \textit{random} baseline.

%% file: Sections/6-Conclusion.tex
\section{Conclusion \& Future Work}\label{sec:conclusion}
In this paper we proposed two methods of using evolutionary algorithms to create QUBO representations of \msat problems automatically. As our first method, we proposed an evolutionary algorithm for creating Pattern QUBOs for any \tsat{} clause in given value ranges. Especially for the commonly used smaller ranges we can quickly and efficiently find valid patterns. Furthermore, we showed how to adapt the fitness function of this method such that the method specifically creates Pattern QUBOs that fulfill certain design criteria, like a user-specified sparsity or a specific energy gap between correct and incorrect assignments. Then, assuming a set of Pattern QUBOs for each type of \tsat{} clause, our second evolutionary algorithm can create full QUBO representations of \msat problems, by selectively choosing a Pattern QUBO from the given set of Pattern QUBOs, one for each clause of the \tsat{} problem individually. This is in contrast to currently known procedures of transforming \msat instances to instances of QUBO, where each type of clause (i.e., zero negations, one negation,...) of the \msat problem is transformed to an instance of QUBO by using a single, predefined Pattern QUBO. This method of assembling QUBOs outperforms our baselines for smaller 500-clause datasets, draws or improves on the performance for longer, more difficult 1000-clause problems, and even shows competitive performance to \sota{} baselines when evaluated on real quantum hardware, highlighting potential for transferring classically optimized (tabu-search) populations to noisy quantum optimizers.

In the future, we aim to delve into optimizing evolutionary strategy parameters, such as population sizes, mutation rates, and different fitness functions to enhance the QUBO generation process for \msat problems. We will also investigate why our evolutionary method generates significantly more Pattern QUBOs for type 3 clauses than for any other type of clause.

In this paper, we trained the second evolutionary algorithm via the classical tabu search method. In the future, we want to train our second evolutionary method directly on D-Wave's quantum annealer to guide the evolutionary search towards creating QUBO representations of \msat problems that perform well when solved on the quantum annealer. Finally, by choosing Pattern QUBOs for each individual clause of a \tsat{} formula, it may be possible to create QUBO representations of \msat problems that possess a certain sparsity or a (more) uniform value range (i.e., the maximum and minimum value of the QUBO entries are not too far apart). We will evaluate adding these criteria as an extension to the fitness function of our second evolutionary method.

%% file: Sections/9-Acknowledgements.tex
\section{Acknowledgments}
The partial funding of this paper by the German Federal Ministry of Education
and Research through the funding program “quantum technologies — from basic
research to market” (contract number: 13N16196) is gratefully acknowledged.
Furthermore, this paper was also partially funded by the German Federal Ministry for Economic Affairs and Climate Action through the funding program "Quantum Computing -- Applications for the industry" (contract number: 01MQ22008A).